\documentclass[pre,showpacs,eqsecnum]{revtex4}
\usepackage{graphicx}
\begin{document}
\title{
Exact solution of a linear molecular motor model driven by
two-step fluctuations and subject to protein friction}
\author{Hans C. Fogedby}
\email{fogedby@phys.au.dk} \affiliation { Institute of Physics and
Astronomy,
University of Aarhus, DK-8000, Aarhus C, Denmark\\
and\\
NORDITA, Blegdamsvej 17, DK-2100, Copenhagen {\O}, Denmark }
\author{Ralf Metzler}
\email{metz@nordita.dk} \affiliation { NORDITA, Blegdamsvej
17,\\ DK-2100, Copenhagen {\O}, Denmark }
\author{Axel Svane}
\email{svane@phys.au.dk} \affiliation { Institute of Physics and
Astronomy,
University of Aarhus,\\
 DK-8000, Aarhus C, Denmark
}
\begin{abstract}
We investigate by analytical means the stochastic equations of
motion of a linear molecular motor model based on the concept of
protein friction. Solving the coupled Langevin equations
originally proposed by Mogilner et al. (A. Mogilner et al., Phys.
Lett. {\bf 237}, 297 (1998)), and averaging over both the two-step
internal conformational fluctuations and the thermal noise, we
present explicit, analytical expressions for the average motion
and the velocity-force relationship. Our results allow for a
direct interpretation of details of this motor model which are not
readily accessible from numerical solutions. In particular, we
find that the model is able to predict physiologically reasonable
values for the load-free motor velocity and the motor mobility.
\end{abstract}
\pacs{87.16.Nn, 02.50.-r, 05.40.-a, 87.10.+e}
\maketitle

\section{\label{introduction}Introduction}
There is currently widespread interest in molecular motors, both
from a biochemical-physiological and a physics point of view.
Whereas the former is mostly concerned with the molecular
structure of motors and their structural interplay with the
support on which they move, physicists study the non-equilibrium
transport properties of motors and their physical interactions
with the support, such as load-velocity relations or adhesion
forces between motor and support. Molecular motors, in general,
are energy-consuming, non-equilibrium nanoscale engines, which are
encountered in various dynamical processes on the intra- and
intercellular level \cite{Howard97,Alberts94}; for a recent review
of the more physical aspects see, for instance,
Ref.~\cite{Reimann01}. Such motors are responsible for
intracellular transport of molecules and small vesicles in
eukaryotic cells; they are powering genomic transcription and
translation, cell division (mitosis), or the packaging of viral
DNA into nanoscale transport containers (capsids)
\cite{Woledge85,Stryer88,Darnell90,Abeles92,Leibler94,Simpson00}.
Larger assemblies of motors working in unison are responsible for
the motility of, e.g., bacteria, they play a role in cell growth,
and they are responsible for muscle contraction leading to
macroscopic motion \cite{Howard97,Alberts94,Reimann01,Badoual02}.

Linear motor proteins like myosin, kinesin, dynein, or DNA
helicase or RNA polymerase are driven by the cyclical hydrolysis
of ATP into ADP and inorganic phosphate and  wander along linear,
polar biomolecular tracks such as actin filaments, microtubules,
RNA or DNA. The motion is typically associated with two or
multi-step conformational changes in the motor protein in
interaction with ATP/ADP and the filament/support, and takes place
in a thermal environment subject to viscous forces.

Modern experimental techniques in biology and biophysics, in
particular single biomolecule manipulation by for example optical
tweezers or micro-needles, and single particle tracking methods,
have yielded considerable insight into the mechanism and the
relevant physical scales in molecular motor behavior
\cite{Svoboda94a,Svoboda94b,Kojima97,Higuchi97,Coppin97,
Wang98,Mehta99,Strick01}. The typical size of a molecular motor is
of order 10 - 20 nm, moving with a step size of order 8 nm, e.g.,
kinesin on microtubules, with one ATP molecule hydrolyzed on the
average per step. The velocities of molecular motors range from
nm/sec to $\mu$m/s and the maximum load is of the order of several
pN (e.g., $\sim$6pN for kinesin on microtubules). However, the
latter can reach up to 57pN for the rotating packaging motor of
bacteriophages \cite{Smith01}. The time scale of the chemical
cycle is a few ms and the average energy input from the ATP-ADP
cycle of order 15-20 kT.

A biomolecular motor represents an interesting and ubiquitous
non-equilibrium system operating in the classical regime and is
thus directly amenable to an analysis using standard methods
within non-equilibrium statistical physics. Physical modelling of
molecular motors has thus been studied intensively in recent years
both from the point of view of the fundamental underlying physical
principles and with regard to specific modelling of concrete
motors \cite{Fisher99,Leibler93,Leibler94,Duke96,Magnasco93,
Julicher97,Astumian97,Astumian99,Astumian02,Ambaye99,Ajdari94}.
More recently, the concerted action of multiple motors has been
considered, such as the action of elastically \cite{Vilfan98} and
rigidly \cite{Badoual02} coupled motors, for instance, in muscles
\cite{Vilfan03}. Motors interacting with freely polymerizing
microtubules or actin filaments give rise to rich pattern
formation such as asters \cite{Lee01}, and are responsible for the
formation of the contractile ring emerging during cell division
\cite{Mulvihill01,Pelham02}.

The most common statistical approach to molecular motors is that
of a ratchet model \cite{Julicher97,Reimann01}, mimicking the
periodically alternating energy landscape (given by the
interaction potential with its support) perceived by the motor
during its mechanochemical cycle. Such ratchet models date back to
Smoluchowski \cite{Smoluchowski12} and Feynman \cite{Feynman63},
and Huxley's pioneering work \cite{Huxley57} on motor proteins
actually corresponds to a Brownian ratchet \cite{Reimann01}. We
note that ratchets play a much more general role, and real-space
ratchets may even be used on the microscale for particle
separation \cite{Kettner00,Matthias03}.

An alternative motor model can be based on protein friction
\cite{Tawada91,Leibler93,Jannink96,Brokaw97,Julicher97,Imafuku96}.
This concept relies on the idea that due to the weak chemical
bonds forming between motor protein and the polar actin or
microtubule track, after elimination of the detailed degrees of
freedom, an effective friction $\zeta_p$ builds up between motor
and track. This protein friction $\zeta_p$ acts like a linear
friction if the associated time scale of motion is longer than the
characteristic time of the kinetics of motor-track
bonds. If not,
no protein friction can build up, and the motor is only subject to
the smaller viscous drag $\zeta_v$ due to the environment. The
protein friction is therefore highly non-linear. On the basis of
this scenario, Mogilner et al. \cite{Mogilner98} recently studied
a simple two-step linear molecular motor represented by two
coupled overdamped oscillators driven by a two-step Markov process
alternating between a relaxed and a strained state of the
oscillators and embedded in a thermal environment represented by
additive white noise. The two subprocesses are associated with
internal conformational changes of the motor protein such that one
subprocess is slow, allowing protein friction to be established,
while the other is fast and only subject to solvent friction. By
means of a numerical analysis Mogilner et al. show that the system
acts like a motor and can carry a load. However, unlike the
ratchet models, which operate with an attachment to a periodic
polar protein filament, the model of Mogilner et al. only needs a
`passive' groove in order to perform directed motion, and the
`ratcheting' comes about by assuming the asymmetric internal
velocity fluctuations, which are then rectified by protein
friction. In that sense, it is a robotic model of molecular
motors.

In the present paper we reanalyze the motor model by Mogilner et
al. from a purely analytical point of view and derive explicit
expressions for the motion of the motor and the velocity-load
relationship. Using the biological parameter values quoted by
Mogilner et al. we show that the model gives rise to
physiologically reasonable values for the motor velocity, whereas
our analysis leads to a correction of the maximum load force by an
order of magnitude in comparison with the numerical results
obtained in Ref.~\cite{Mogilner98}. This discrepancy is associated
with a difference in the dynamics of the analytical model as
compared with the numerical simulation. Allowing for a larger
relaxation rate the analytical result for the maximum load force
approaches the biological regime.
The paper is organized in the following manner. In
Sec.~\ref{model} we introduce the model. In Sec.~\ref{analysis} we
solve the model analytically. In Sec.~\ref{discussion} we discuss
the results and compare with Ref.~\cite{Mogilner98}. The paper
ends with a summary and a conclusion in Sec.~\ref{conclusion}
\section{\label{model}Model}
\begin{figure}
\includegraphics[width=0.5\hsize]{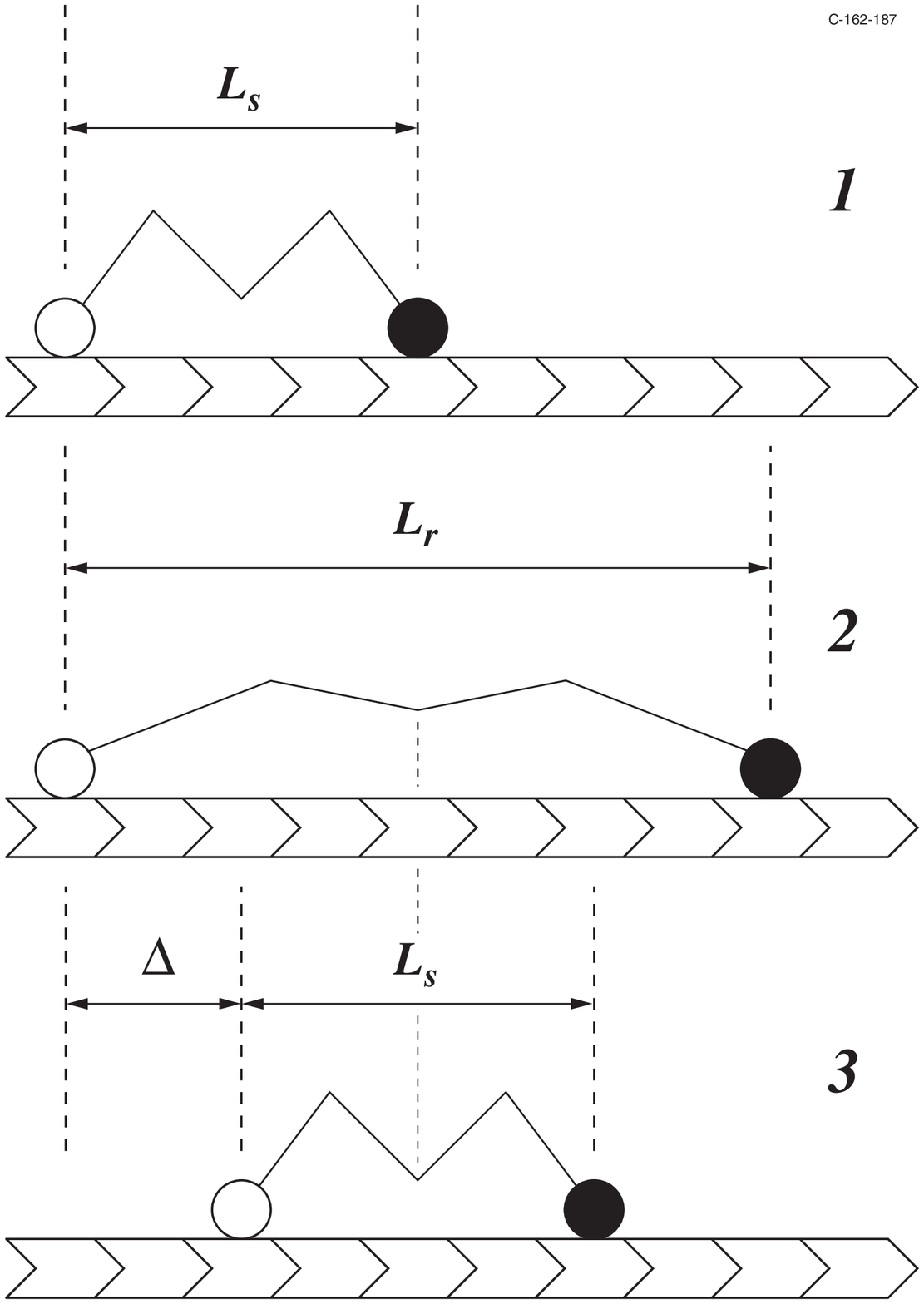}
\caption{Molecular motor model showing one mechanochemical cycle,
during which internal fluctuations become directed (`ratcheted')
through protein friction. The conformational changes of the motor
protein are represented by two states of an effective spring with
rest lengths $L_s$ for the strained and $L_r$ for the relaxed
states. $1\to 2$: Slow relaxation of the previously strained
spring to assume the rest length $L_r$; due to protein friction,
the white working head of the motor stays attached to the polar
biopolymer (actin filament or microtubule), while the black idle
head is free to move. $2\to 3$: As a consequence of ATP hydrolysis
(`power stroke'), the spring contracts so quickly that the protein
friction breaks down, and both heads symmetrically converge to
assume the strained configuration with rest length $L_s$. The
distance covered per mechanochemical cycle is $\Delta=
(L_r-L_s)/2$. (Adopted from Ref.~\protect\cite{Mogilner98}.)}
\label{fig1}
\end{figure}

The motor model based on protein friction which was introduced in
Ref.~\cite{Mogilner98}, is defined as follows, compare
Fig.~\ref{fig1}. Assume that the mechanochemical cycle of the
motor protein walking along the track made up of an actin filament
or microtubule can be pinned down to the periodical switching
between two states, and that each of these two states can be
described by two motor heads connected by an effective spring
representing the backbone of the motor protein between these
heads. From the strained state (S), characterized by a rest length
$L_s$, the motor protein converges towards a relaxed state (R)
with rest length $L_r>L_s$, i.e., the distance between the motor
heads increases. This process is slow enough to make sure that the
adhesion between the motor's `working head' (white circle in
Fig.~\ref{fig1}) and track stays intact, in such a manner that
asymmetric motion with respect to the track is achieved (stick).
In contrast, during the fast `power stroke' from the relaxed state
back to the strained state after hydrolysis of ATP, the protein
friction is broken and both heads move under the low Reynolds
number conditions of the environment, such that both heads
symmetrically approach each other and assume the rest length $L_s$
(slip).

This model can be cast into the two coupled Langevin equations:
\begin{eqnarray}
\zeta(t)\dot x(t) &=& -f+k(t)\frac{y(t)-x(t)-L(t)}{2} + N_x(t),
\label{lan1}
\\
\zeta_v\dot y(t) &=& -k(t)\frac{y(t)-x(t)-L(t)}{2} + N_y(t),
\label{lan2}
\end{eqnarray}
in which we have introduced the time-dependent friction
coefficient $\zeta(t)$ in comparison to Ref.~\cite{Mogilner98} for
convenience, to account for the cyclical attachment to the track.
In Eqs.~(\ref{lan1},\ref{lan2}) the variable $x$ and $y$ represent
the positions of the two heads of the motor molecule along the
track, corresponding to the equations of motion of two coupled,
overdamped oscillators. The coordinate of the idle head $y$ is
associated with a viscous friction drag coefficient $\zeta_v$ of
order $6 \pi\eta r$ (Stokes law), where $\eta$ is the viscosity of
water and $r$ the size of the motor protein head. The same
friction acts on the working head during the fast conformational
change R$\to$S, whereas during the slow process S$\to$R, it
experiences the protein friction drag with coefficient $\zeta_p$
\cite{Tawada91,Leibler93,Jannink96,Brokaw97,Julicher97,Imafuku96},
corresponding to a stick-slip motion of the working head. The
model equations (\ref{lan1},\ref{lan2}) are driven by thermal
noises $N_x(t)$ and $N_y(t)$, with $\langle N_{x,y}(t)\rangle =0$,
representing the ambient environment with correlations
\begin{eqnarray}
\langle N_{x,y}(t)N_{x,y}(t')\rangle=
2k_{\text{B}}T\zeta_{p,v}\delta(t-t'), \label{noise}
\end{eqnarray}
balancing the friction terms by means of the
fluctuation-dissipation theorem. We note that during the detached
straining step, the role of working and idle heads may be
exchanged (i.e., the motor heads may be turned around a common
axis), as was recently demonstrated for kinesin motor heads
\cite{Kaseda03}.

The conformational changes of the motor driven by the ATP-ADP
hydrolytic cycle and the cyclical attachment to the substrate
correspond to a continuous two-state Markov process for the time
dependent rest length $L(t)$, the time dependent spring constant
$k(t)$, and the time dependent protein friction $\zeta(t)$,
alternating between state R with rest length $L_r$, spring
constant $k_r$, and protein friction $\zeta=\zeta_p$ and state S
with rest length $L_s$, spring constant $k_s$, and viscous
friction $\zeta=\zeta_v$. The power stroke conformational
transition R$\to$S driven by the ATP hydrolysis is characterized
by the rate $g_s$; the relaxational conformation change S$\to$R
has the rate $g_r$.

The relevant biological parameters quoted in
Ref.~\cite{Mogilner98}, entering Eqs. (\ref{lan1}) and
(\ref{lan2}), are: Rate of hydrolysis $g_s\sim 10^3$~s$^{-1}$,
relaxation rate $g_r\sim 10^3$~s$^{-1}$, spring coefficient in
relaxed state $k_r\sim 0.01$~pN/nm, spring coefficient in strained
state $k_s\sim 0.5$~pN/nm, rest length in relaxed state $L_r\sim
40$~nm, rest length in strained state $L_s\sim 20$~nm, viscous
drag coefficient $\zeta_v\sim 10^{-6}$~pN~s/nm, protein friction
drag coefficient $\zeta_p\sim 5\times 10^{-5}$~pN~s/nm, and load
force $f\sim \pm 1$~pN. For further discussion of the model and
parameter choices under biological conditions we refer to ref.
\cite{Mogilner98}. An important difference between the model of
Mogilner at al. and the present one in Eqs. (\ref{lan1}) and
(\ref{lan2}) is that we take $\zeta(t)=\zeta_v$ during the entire
duration of the strained state $S$, while Mogilner assumes
$\zeta(t)=\zeta_v$ only in the first short time interval as the
spring contracts (a time interval of order
$t_{\text{slip}}\sim\zeta_v/k_s$; in the simulations of
Ref.~\cite{Mogilner98} $t_{\text{slip}}$ is taken infinitesimally
small), after which the protein bonds will form and protein
friction take over, $\zeta(t)=\zeta_p$. The two models will be
similar if the relaxation time $g_r^{-1}$ is of the order of
$t_{\text{slip}}$.
\section{\label{analysis}Analysis}
In this Section, we present a solution scheme for this motor
model. The  results obtained are then further analyzed in the
following Section.

\subsection{Analytical solution}
The motor equations (\ref{lan1}) and (\ref{lan2}) are readily
analyzed by (i) solving Eq. (\ref{lan1}) for $y(t)$ and deriving
$dy/dt$, (ii) eliminating $y$ in Eq. (\ref{lan2}) and setting the
two expressions equal to one another. We thus obtain the following
equations for  $v_x=dx/dt$ and $v_y=dy/dt$:
\begin{eqnarray}
&&2(\zeta\zeta_v/k)\dot v_x+ (\zeta+ \zeta_v-2\zeta\zeta_v(\dot
k/k^2)+2\dot\zeta\zeta_v/k)v_x = \nonumber \label{lan11}
\\
&&-f(1-2\zeta_v(\dot k/k^2))-\zeta_v\dot L + 2\zeta_v\dot N_x/k +
(1-2\zeta_v\dot k/k^2)N_x+N_y,
\\
&&\zeta_v v_y=-f -\zeta v_x +N_x+N_y. \label{lan22}
\end{eqnarray}
Denoting the initial velocity at time $t=0$ by $v_{x0}$ Eq.
(\ref{lan11}) is readily solved by quadrature and together with
Eq. (\ref{lan22}) we obtain
\begin{eqnarray}
&&v_x(t)=\frac{k(t)e^{-\gamma(t)}}{\zeta(t)} \left[
v_{x0}\frac{\zeta(0)}{k(0)}-\frac{1}{2\zeta_v} \int_0^t dt'[\tilde
f(t')+\zeta_v\dot L(t') -\tilde N(t')]e^{\gamma(t')} \right],
\label{sol1}
\\
&&v_y(t)=-\frac{f}{\zeta_v} -
\frac{\zeta(t)}{\zeta_v}v_x(t)+\frac{1}{\zeta_v}[N_x(t)+N_y(t)]
\label{sol2}
\end{eqnarray}
which form the basis for our discussion.

We have introduced the renormalized load force $\tilde f$, the
renormalized noise $\tilde N$, and the integrated spring and
friction constant $\gamma$:
\begin{eqnarray}
&&\tilde f =f(1-2\zeta_v(\dot k/k^2)), \label{frenorm}
\\
&&\tilde N =2\zeta_v\dot N_x/k + (1-2\zeta_v\dot k/k^2)N_x+N_y,
\\
&&\gamma(t)=(1/2\zeta_v)\int_0^t k(t')dt' + \int_0^t
k(t')/2\zeta(t')dt' \label{gamma} .
\end{eqnarray}
\subsection{General properties}
We note various general features of this solution. First, both the
load force $f$ and the thermal noises $N_x$ and $N_y$ are
renormalized by the fluctuating spring constant $k$. Secondly, the
thermal noise basically enters additively and entails thermal
fluctuations of the velocities. Since the stochastic
conformational changes giving rise to the fluctuations of $k$,
$L$, and $\zeta$ are independent of the thermal fluctuations we
can in order to monitor the time dependence of the mean motor
velocity with impunity average over the noise. Note that the heat
bath, of course, still enters through the friction coefficients.
In the long time steady state limit we can ignore the initial
terms and we obtain the reduced equations for the thermally
averaged velocities
\begin{eqnarray}
&&v_x(t)=-\frac{k(t)e^{-\gamma(t)}}{2\zeta_v\zeta(t)}\int_0^t
dt'[\tilde f(t')+\zeta_v\dot L(t')]e^{\gamma(t')}, \label{sol11}
\\
&&v_y(t)= +\frac{k(t)e^{-\gamma(t)}}{2\zeta_v^2}\int_0^t
dt'[\tilde f(t')+\zeta_v\dot
L(t')]e^{\gamma(t')}-\frac{f}{\zeta_v} \label{sol22},
\end{eqnarray}
which we proceed to discuss.
\subsection{Constant spring constant and rest length}
Let us first as an illustration consider the case of a constant
spring length $L$, a constant spring constant $k$, and a constant
protein friction $\zeta(t)=\zeta_p$. In this simple case
$\gamma(t) = k((\zeta_p+\zeta_v)/2\zeta_p\zeta_v))t$ and the load
force is unrenormalized. We obtain
\begin{eqnarray}
v_x = v_y= -\frac{f}{\zeta_p+\zeta_v}.
\end{eqnarray}
Here the load $f$ after a transient period drives the idle and
working heads with a constant mean velocity. Defining the mobility
according to
\begin{eqnarray}
v_x= -\mu f,
\label{mobdef}
\end{eqnarray}
we infer the mobility in the absence of conformational
fluctuations
\begin{eqnarray}
\mu = \frac{1}{\zeta_p+\zeta_v}.
\label{mob}
\end{eqnarray}
In the absence of a load for $f=0$, the mean velocity vanishes and
the system does not move, i.e., we do not have motor properties.
This is also a statement of the second law of thermodynamics
expressing the fact that we cannot extract work from a system in
thermal equilibrium. In the case of constant $k$, constant $L$,
and constant $\zeta$ the coupled Langevin equations describe the
temporal fluctuations of a system in thermal equilibrium. The
motor property is thus necessarily due to the fluctuating spring
constant $k(t)$ and fluctuating length $L(t)$ characterizing the
conformational fluctuations in combination with the cyclical
attachment described by the fluctuating friction coefficient
$\zeta(t)$.
\subsection{Fluctuating spring constant,
rest length, and protein friction}
The idea behind the model is that fluctuations of the spring
constant $k(t) =k_s,k_r$ ($k_r<k_s$) and rest length
$L(t)=L_s,L_r$ ($L_r>L_s$), modelling the ATD-ADP driven
conformational changes, provide an energy source. In combination
with the synchronized stick-slip mechanism modelled by a
fluctuating protein friction $\zeta(t)=\zeta_p,\zeta_v$ ($\zeta_p>
\zeta_v$), this process can drive the system in the absence of a
force. This mechanism is modelled by the two-step Markovian
process S$\leftrightarrow$R with relaxation rates $g_r$ for
S$\to$R and $g_s$ for R$\to$S. The master equations for this
process denoting the corresponding probabilities by $P_s(t)$ and
$P_r(t)$ thus take the form:
\begin{eqnarray}
&&\frac{dP_s(t)}{dt}=g_sP_r(t)-g_rP_s(t) \label{mas1},
\\
&&\frac{dP_r(t)}{dt}=g_rP_s(t)-g_sP_r(t) \label{mas2},
\end{eqnarray}
with stationary solutions
\begin{eqnarray}
&&P_s=\frac{g_s}{g_s+g_r}, \label{ps}
\\
&&P_r=\frac{g_r}{g_s+g_r}. \label{pr}
\end{eqnarray}
The stationary mean value of, e.g., the spring constant, is thus
given by
\begin{eqnarray}
k_{\text{st}}= k_sP_s+k_rP_r = \frac{k_sg_s+k_rg_r}{g_s+g_r}.
\label{kstat}
\end{eqnarray}
For the present purposes it turns out to be more convenient in
discussing the conformational transitions to focus on the
probability distributions $\tilde P_s(t)$ and $\tilde P_r(t)$
characterizing the residence of the system in either the strained
state or the relaxed state at a time $t$. The distribution is
exponential in time and we obtain properly normalized
\begin{eqnarray}
&&\tilde P_s(t)=g_r e^{-g_rt},
\\
&&\tilde P_r(t)=g_s e^{-g_st}.
\end{eqnarray}
The mean values of the residence times are then given by
\begin{eqnarray}
\label{residence}
&&\langle t\rangle_s=\frac{1}{g_r},
\\
&&\langle t\rangle_r=\frac{1}{g_s}.
\end{eqnarray}
The mean value of $k$ may then be obtained as a time average:
\begin{eqnarray}
\langle k\rangle = \frac{k_s\langle t\rangle_s + k_r\langle
t\rangle_r}{\langle t\rangle_s+\langle t\rangle_r}=
\frac{k_s/g_r+k_r/g_s}{1/g_r+1/g_s},
\end{eqnarray}
in accordance with Eq. (\ref{kstat}).
\subsection{The motor property without a load}
Here we establish the fundamental motor property of the model in
the absence of a load. For $f=0$ we have from Eqs. (\ref{sol11})
and (\ref{sol22}),
\begin{eqnarray}
&&v_x^0=-\frac{k(t)e^{-\gamma(t)}}{2\zeta(t)}\int_0^tdt'\dot
L(t')e^{\gamma(t')}, \label{sol111}
\\
&&v_y^0=+\frac{k(t)e^{-\gamma(t)}}{2\zeta_v}\int_0^tdt'\dot
L(t')e^{\gamma(t')}, \label{sol222}
\end{eqnarray}
At a superficial glance it looks like the motor heads move in
opposite direction. However, subtracting the velocities and noting
that
\begin{eqnarray}
\dot\gamma(t) = \frac{k(t)}{2\zeta(t)}+\frac{k(t)}{2\zeta_v},
\end{eqnarray}
we obtain
\begin{eqnarray}
v_x^0-v_y^0=-\dot\gamma(t)e^{-\gamma(t)}\int_0^t~dt'\dot
L(t')e^{\gamma(t')}.
\end{eqnarray}
Finally, assuming ergodicity (to be established later) and time
averaging in combination with partial integrations we have for
$T\rightarrow\infty$
\begin{eqnarray}
\langle v_x^0\rangle - \langle v_y^0\rangle &=&
-\frac{1}{T}\int_0^Tdt\dot\gamma(t)e^{-\gamma(t)}\int_0^tdt'\dot
L(t')e^{\gamma(t')} \nonumber
\\
&=&\frac{1}{T}\int_0^Tdt\frac{d}{dt}\left[e^{-\gamma(t)}\right]
\int_0^tdt'\dot L(t')e^{\gamma(t')} \nonumber
\\
&=&-\frac{1}{T}\int_0^T\dot L(t)dt = 0 \label{motorproperty},
\end{eqnarray}
where the last step corresponds to an integration by parts (note that
$\gamma(t)$ is monotonically increasing, and therefore $T^{-1}\left[\int_0^t
\dot{L}(t')\exp\{\gamma(t')-\gamma(t)\}\right]_0^T<T^{-1}\left[L(t)\right]
_0^T\to 0$),
and we conclude that the average velocities of the two heads are
in fact identical: The working head and the idle head move
together.

Next we derive an explicit expression for $\langle v_x^0\rangle$.
Introducing the auxiliary fluctuating variable
\begin{eqnarray}
a(t)=\frac{k(t)}{2}\left[\frac{1}{\zeta_v}-\frac{1}{\zeta(t)}\right],
\label{aux}
\end{eqnarray}
and using the above result we obtain by adding Eqs. (\ref{sol111})
and (\ref{sol222}):
\begin{eqnarray}
\langle v_x^0\rangle=\frac{1}{2}\int_0^t dt'\langle
e^{-(\gamma(t)-\gamma(t'))}a(t)\dot L(t')\rangle. \label{vel0}
\end{eqnarray}
Here $\langle\cdots\rangle$ denotes an ensemble average with
respect to the conformational fluctuations. Since $L(t)$, $k(t)$,
and $\zeta(t)$ are governed by the same stochastic process,
with the values $L_s$, $\zeta_v$, $k_s$, and $L_r$, $\zeta_p$, $k_r$ in
the strained and relaxed states, respectively, we obtain
from the expressions for $a$ and $\dot\gamma$ in Eqs. (\ref{aux})
and (\ref{gamma})
\begin{eqnarray}
&&a_r=\frac{k_r}{2}\left[\frac{1}{\zeta_v}-\frac{1}{\zeta_p}\right],
\label{ar}
\\
&&a_s=0,
\label{as}
\\
&&\dot\gamma_r=
\frac{k_r}{2}\left[\frac{1}{\zeta_v}+\frac{1}{\zeta_p}\right],
\label{gammar}
\\
&&\dot\gamma_s=\frac{k_s}{\zeta_v} \label{gammas}.
\end{eqnarray}
Denoting the jump times for the transitions between the strained
and relaxed state by  $t_n$, $n=1, 2, \cdots$ and assuming that
the system is in a relaxed state at $0<t<t_1$ we have
\begin{eqnarray}
\dot L(t)=(L_r-L_s)\sum_{n=1}^\infty (-1)^n \delta(t-t_n),
\end{eqnarray}
and inserting in Eq. (\ref{vel0}) we obtain
\begin{eqnarray}
\langle v_x^0\rangle= \frac{L_r-L_s}{2} \sum_{n=1}^N(-1)^n
\left\langle a(t)\exp\left({-\int_{t_n}^t\dot\gamma(t')dt'}\right)
\right\rangle, \label{int11}
\end{eqnarray}
where $t_N<t<t_{N+1}$. Note that the exponential term is just a more
complicated way to write $\exp\{-\gamma(t)+\gamma(t_n)\}$, which will
be useful below. For even $N$ the system is in the relaxed
state for $t_N<t'<t$ with probability $P_r$. Similarly for odd $N$
the motor ends in the strained  state which occurs with
probability $P_s$. Introducing the time interval
$\tau_n=t_{n+1}-t_n$, noting that the residence distributions are
statistically independent, and introducing the notation
\begin{eqnarray}
&&R=\langle\exp{(-\dot\gamma_r \tau)}\rangle_r=\int_0^\infty
d\tau\tilde P_r(\tau)\exp{(-\dot\gamma_r
\tau)}=\frac{g_s}{g_s+\dot\gamma_r},
\label{r}
\\
&&S=\langle\exp{(-\dot\gamma_s \tau)}\rangle_s=\int_0^\infty
d\tau\tilde P_s(\tau)\exp{(-\dot\gamma_s
\tau)}=\frac{g_r}{g_r+\dot\gamma_s},
\label{s}
\end{eqnarray}
the mean velocity can be expressed in terms of geometrical series,
\begin{eqnarray}
\langle v_x^0\rangle= \frac{L_r-L_s}{2}\left[P_r
a_rR(1-S)\sum_{n=0}(SR)^{n}-P_s a_s
S(1-R)\sum_{n=0}(SR)^{n}\right],
\end{eqnarray}
or summing the series (completing $N\to\infty$),
\begin{eqnarray}
\langle v_x^0\rangle=\frac{L_r-L_s}{2}\frac{P_r a_rR(1-S)-P_s a_s
S(1-R)}{1-RS}.
\end{eqnarray}
Inserting $P_r$, $P_s$, $R$, and $S$ from Eqs. (\ref{pr}),
(\ref{ps}), (\ref{r}), and (\ref{s}) we arrive at
\begin{eqnarray}
\langle v_x^0\rangle= \frac{L_r-L_s}{2}
\frac{g_rg_s(a_r\dot\gamma_s-a_s\dot\gamma_r)}
{(g_s\dot\gamma_s+g_r\dot\gamma_r+\dot\gamma_r\dot\gamma_s)(g_s+g_r)}.
\label{int22}
\end{eqnarray}
First we note that the expression vanishes for
$a(t)=\dot\gamma(t)$ thus corroborating the validity of the time
average in Eq. (\ref{motorproperty}) and establishing ergodicity.
Finally, inserting $a_r$, $a_s$, $\dot\gamma_r$, and $\dot\gamma_s$
from Eqs. (\ref{ar}), (\ref{as}), (\ref{gammar}), and
(\ref{gammas}) we obtain for the explicit expression for the motor
velocity in the absence of a load
\begin{eqnarray}
\langle v_x^0\rangle= \frac{(L_r-L_s)(\zeta_p -\zeta_v)
k_rk_sg_rg_s}{2(g_r+g_s)[2g_sk_s\zeta_p\zeta_v+
k_r(\zeta_p+\zeta_v)(g_r\zeta_v+k_s)]}.
\label{noloadv}
\end{eqnarray}
%
\subsection{The motor property with load}
We next turn to the case of a load force applied to the motor.
First we establish that in the presence of the load the two heads
of the motor move together with the same average velocity. From
Eqs. (\ref{sol11}) and (\ref{sol22}) we obtain
\begin{eqnarray}
v_x-v_y =v_x^0-v_y^0
-\frac{\dot\gamma(t)e^{-\gamma(t)}}{\zeta_v}\int_0^tdt'\tilde
f(t')e^{\gamma(t')} + \frac{f}{\zeta_v}.
\end{eqnarray}
Inserting $\tilde f$ from Eq. (\ref {frenorm}) and averaging over
time we have, using Eq. (\ref{motorproperty}):
\begin{eqnarray}
\langle v_x\rangle-\langle v_y\rangle =&&
-\frac{f}{\zeta_v}\frac{1}{T}\int_0^Tdte^{-\gamma(t)}\dot\gamma(t)
\int_0^tdt' e^{\gamma(t')} + \frac{f}{\zeta_v} \nonumber
\\
&&+f\frac{2}{T}\int_0^Tdt
e^{-\gamma(t)}\dot\gamma(t)\int_0^tdt'\frac{\dot k(t')}{k(t')^2}
e^{\gamma(t')}.
\label{vdif}
\end{eqnarray}
Performing the integrals by partial integration along the same
lines as in the load-free case, the first two terms in Eq.
(\ref{vdif}) cancel and we find in the limit $T\to\infty$,
\begin{eqnarray}
\langle v_x\rangle=\langle v_y\rangle,
\end{eqnarray}
i.e., the two motor heads move together with the same average
velocity.

We now turn to the evaluation of the load-velocity relationship.
From Eqs. (\ref{sol11}) and (\ref{sol22}) and inserting $\tilde f$
we have
\begin{eqnarray}
&&v_x=v_x^0-\frac{f}{2\zeta_v}\int_0^tdt' \frac{k(t)}{\zeta(t)}
\left[1+2\zeta_v\frac{d}{dt'}\left(\frac{1}{k(t')}\right)\right]
e^{-(\gamma(t)-\gamma(t'))},
\label{fvelo1}
\\
&&v_y=v_y^0-\frac{f}{\zeta_v}+\frac{f}{2\zeta_v}\int_0^tdt'
\frac{k(t)}{\zeta_v}
\left[1+2\zeta_v\frac{d}{dt'}\left(\frac{1}{k(t')}\right)\right]
e^{-(\gamma(t)-\gamma(t'))}.
\label{fvelo2}
\end{eqnarray}
From the synchronization of the stochastic processes we obtain the
identity
\begin{equation}
\frac{1}{k(t)}=\frac{1}{k_r}+\frac{1/k_s-1/k_r}
{\dot\gamma_s-\dot\gamma_r}(\dot\gamma(t)-\dot\gamma_r),
\end{equation}
and therefore
\begin{equation}
\frac{d}{dt}\frac{1}{k(t)}=\frac{1/k_s-1/k_r}
{\dot\gamma_s-\dot\gamma_r}\ddot\gamma(t).
\end{equation}
Inserting into Eqs.~(\ref{fvelo1}) and (\ref{fvelo2}) and
averaging
\begin{eqnarray}
\langle v_x\rangle=\langle
v_x^0\rangle&-&\frac{f}{2\zeta_v}\left\langle\frac{k(t)}
{\zeta(t)}\int_0^t dt' e^{-(\gamma(t)-\gamma(t'))}\right\rangle
\nonumber
\\
&-&f\frac{1/k_s-1/k_r}{\dot\gamma_s-\dot\gamma_r}\left\langle\frac{k(t)}
{\zeta(t)}\int_0^t dt'
e^{-(\gamma(t)-\gamma(t'))}\ddot\gamma(t')\right\rangle,
\label{int1}
\\
\langle v_y\rangle=\langle v_y^0\rangle &-& \frac{f}{\zeta_v}
+\frac{f}{2\zeta_v}\left\langle\frac{k(t)}{\zeta_v}\int_0^t dt'
e^{-(\gamma(t)-\gamma(t'))}\right\rangle \nonumber
\\
&+&f\frac{1/k_s-1/k_r}{\dot\gamma_s-\dot\gamma_r}\left\langle\frac{k(t)}
{\zeta_v}\int_0^t dt'
e^{-(\gamma(t)-\gamma(t'))}\ddot\gamma(t')\right\rangle.
\label{int2}
\end{eqnarray}
The first integral in Eqs. (\ref{int1}) and (\ref{int2}) has the
form
\begin{eqnarray}
I_1=\left\langle b(t)\int_0^t dt
e^{-(\gamma(t)-\gamma(t'))}\right\rangle =\left\langle
b(t)\int_0^t dt \exp\left(-\int_{t'}^t
dt''\dot\gamma(t'')\right)\right\rangle,
\end{eqnarray}
and is performed by breaking up the integration over $\dot\gamma$
in the exponents and averaging over the time segments yielding
again a geometrical series in terms of  $SR$. We obtain as an
intermediate result
\begin{eqnarray}
I_1=
P_rb_r\left(\frac{1-R}{\dot\gamma_r}+\frac{R(1-S)}{\dot\gamma_s}\right)
\sum_{n=0}(SR)^n
+P_sb_s\left(\frac{1-S}{\dot\gamma_s}+\frac{S(1-R)}{\dot\gamma_r}\right)
\sum_{n=0}(SR)^n,
\end{eqnarray}
and performing the sum and inserting $R$ and $S$ from Eqs.
(\ref{r}) and (\ref{s}),
\begin{eqnarray}
I_1=
\frac{b_rP_r(g_r+g_s+\dot\gamma_s)+b_sP_s(g_r+g_s+\dot\gamma_r)}
{g_s\dot\gamma_s+g_r\dot\gamma_r+\dot\gamma_r\dot\gamma_s}.
\end{eqnarray}
The second integral has the structure
\begin{eqnarray}
I_2=\left\langle c(t)\int_0^t dt'
e^{-(\gamma(t)-\gamma(t'))}\ddot\gamma(t')\right\rangle,
\end{eqnarray}
and was performed in the load-free case in Eqs. (\ref{int11}) to
(\ref{int22}). We found
\begin{eqnarray}
I_2= - \frac{(\dot\gamma_s-\dot\gamma_r)P_rP_s(g_s+g_r)
(c_r\dot\gamma_s-c_s\dot\gamma_r)}
{g_s\dot\gamma_s+g_r\dot\gamma_r+\dot\gamma_r\dot\gamma_s}.
\end{eqnarray}
It is again convenient to introduce the mobility $\mu$ according
to the relation
\begin{eqnarray}
\langle v_x\rangle = \langle v_x^0\rangle -\mu f,
\end{eqnarray}
and we obtain inserting $c(t)=b(t)=k(t)/\zeta_p(t)$ for $\langle
v_x\rangle$, or $c(t)=b(t)=k(t)/\zeta_v$ for $\langle v_y\rangle$
\begin{eqnarray}
\mu = &+&\frac{(k_r/2\zeta_v\zeta_p)P_r(g_r+g_s+\dot\gamma_s)+
(k_s/2\zeta_v^2) P_s(g_r+g_s+\dot\gamma_r)}
{g_s\dot\gamma_s+g_r\dot\gamma_r+\dot\gamma_r\dot\gamma_s}
\nonumber
\\
&-&\frac{P_rP_s(g_s+g_r)((k_r/\zeta_p)\dot\gamma_s-
(k_s/\zeta_v)\dot\gamma_r)(1/k_s-1/k_r)}
{g_s\dot\gamma_s+g_r\dot\gamma_r+\dot\gamma_r\dot\gamma_s}.
\label{loadv}
\end{eqnarray}
%
\section{\label{discussion}Discussion}
In this Section, we examine more closely our results derived above,
and compare them to the analysis in Ref.~\cite{Mogilner98}.

\subsection{Free motor}
Let us first examine the simple motor properties in the absence of
a cargo, i.e., for $f=0$. Here the expression in Eq.~(\ref{noloadv})
is at variance with the heuristic expression given
by Mogilner \cite{Mogilner98},
\begin{eqnarray}
\label{mogil}
\langle
v_x\rangle^0_{\text{M}}=\frac{g_sg_r}{g_r+g_s}\frac{L_r-L_s}{2},
\end{eqnarray}
which is solely based on the reaction rates, neglecting the
internal dynamics of the motor. To compare the expression for the
present model, Eq.~(\ref{noloadv}) we introduce the dimensionless
parameters
\begin{eqnarray}
\label{qs} q_s=\frac{g_r}{\dot\gamma_s},
\end{eqnarray}
and
\begin{eqnarray}
\label{qr}
q_r=\frac{g_s}{\dot\gamma_r},
\end{eqnarray}
which express the ratio between the spring relaxation times,
$\dot\gamma_s^{-1}$ and $\dot\gamma_r^{-1}$, and the residence
times Eqs. (\ref{residence}) in states $S$ and $R$, respectively.
In terms of these parameters we obtain
\begin{eqnarray}
\langle v_x^0\rangle= \langle
v_x\rangle^0_{\text{M}}\times\frac{\zeta_p-\zeta_v}{\zeta_p+\zeta_v}
\frac{1}{1+q_s+q_r}.
\label{noloadv2}
\end{eqnarray}
The correction factor to the heuristic velocity given by Mogilner et al.
is clearly smaller than 1, but approaches 1 in the limit of $\zeta_p\gg \zeta_v$
and $q_s\ll 1$, $q_r\ll 1$, which are exactly the conditions under which
expression (\ref{mogil}) was derived.

We note that the velocity vanishes for $\zeta_p=\zeta_v$. In this
case the attachment to the track has no effect on the friction and
there is no motion. In the limit of large protein friction
compared to the viscous drag coefficient, $\zeta_p\gg\zeta_v$, but
$q_s\sim 1$ and/or $q_r\sim 1$ each conformational cycle does not
yield a full step of length $\Delta L/2$ due to incomplete spring
relaxation, and the average velocity is reduced. In the limit of
either $q_s\gg 1$ or $q_r\gg 1 $ the motor comes to rest, as the
relaxation rate or the hydrolysis rate become too large for the
spring to change its average length. The motor would also function
under conditions $\zeta_v > \zeta_p$ or $L_s >L_r$, it would just
move in the opposite direction.

Inserting the characteristic biological numbers from
Ref.~\cite{Mogilner98} we have $\zeta_p/\zeta_v\approx 50$,
$q_r\approx 0.2$, and $q_s\approx 2\cdot 10^{-3}$, and the
correction factor takes a value of about $0.8$. This corresponds
to a average velocity of $\langle v_x\rangle^0\sim 4\times
10^3~\text{nm/s}$. However, under different conditions, the
discrepancy between the heuristic result (\ref{mogil}) and the
exact quantity (\ref{noloadv2}) may become more significant.

\begin{figure}
\includegraphics[width=0.5\hsize,angle=-90]{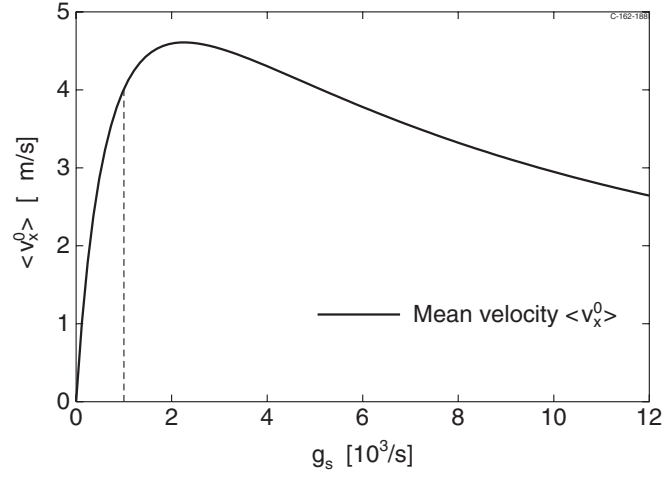}
\caption{Dependence of the mean velocity $\langle v_x^0\rangle$ on
the hydrolysis rate $g_s$, exhibiting a maximum at around
$2.25\cdot 10^3/$s. The vertical dashed line shows the parameter
value given in Ref.~\protect\cite{Mogilner98}. For large values of
$g_s$, the velocity tends to zero. } \label{fig2}
\end{figure}

\begin{figure}
\includegraphics[width=0.5\hsize,angle=-90]{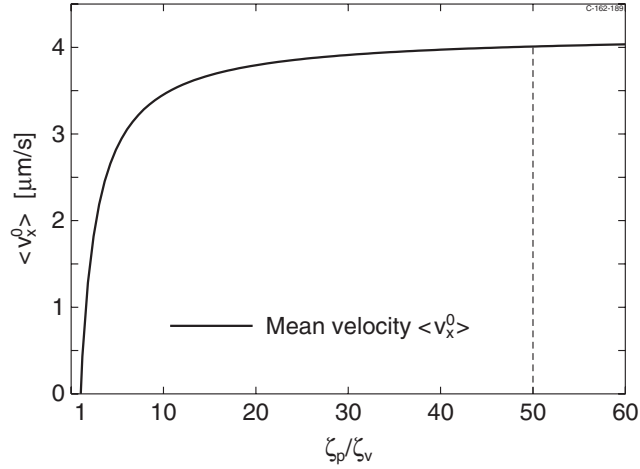}
\caption{Dependence of the mean velocity $\langle v_x^0\rangle$ on
the protein friction $\zeta_p$, reaching a plateau for large
values ($\zeta_v= 10^{-6}$pN s/nm). The protein friction
$\zeta_p=5\times 10^{-5}$nm/(s pN) used in the calculations
corresponds to the dimensionless value 50 in the plot, indicated
by the dashed line. Note that $\langle v_x^0\rangle=0$ corresponds
to $\zeta_p/ \zeta_v=1$.} \label{fig3}
\end{figure}

In Figs.~\ref{fig2} and \ref{fig3}, we show the dependence of the
load-free velocity on the hydrolysis rate $g_s$ and the protein
friction $\zeta_p$ (all other parameters fixed at the values of
Ref.~\cite{Mogilner98}). Accordingly, with respect to these values
the model motor velocity is close to optimum. The maximum in the
$g_s$-dependence shows the interplay between on and off-rates in
the protein friction model, whereas the final plateau in the
$\zeta_p$-dependence indicates the above-mentioned saturation,
i.e., the motor still works for extremely large values of
$\zeta_p$, as long as $q_s$ and/or $q_r$  do not increase to high
values, as well. We note that, as expected, the velocity goes to
zero for vanishing hydrolysis rate, and when the protein friction
approaches $\zeta_p\to\zeta_v$.

\subsection{Motor carrying a load}
In the case of a load or cargo we proceed to discuss the
expression for the motor mobility  in Eq. (\ref{loadv}), which can
be rewritten in the more convenient form
\begin{equation}
\mu=\frac{2(g_r+g_s)(P_rk_r+P_sk_s)(P_r\zeta_v+P_s\zeta_p)+k_rk_s(1+
P_r+P_s\zeta_p/\zeta_v)}{2\zeta_v\left\{2g_sk_s\zeta_p+k_r(g_r\zeta_v+
k_s)(1+\zeta_p/\zeta_v)\right\}}.
\label{loadv2}
\end{equation}
In the further discussion of the mobility it is convenient to
introduce the dimensionless parameters in Eqs. (\ref{qs}) and
(\ref{qr}). The expression (\ref{loadv2}) can then be reduced to
the form
\begin{equation}
\label{loadv3}
\mu=
\frac{(P_r\zeta_v+P_s\zeta_p)(2q_s+(1+\frac{\zeta_v}{\zeta_p})q_r)+(1+P_r)\zeta_v+
P_s\zeta_p }{
2\zeta_v(\zeta_v+\zeta_p) ( 1+q_s+q_r)}.
\end{equation}
Let us investigate this expression in some limiting cases: (i) In
the absence of fluctuations, i.e., the case of a constant spring
constant and rest length, in the relaxed state R, the protein
friction $\zeta(t)$ is locked onto $\zeta_p$, and we have $P_r=1$,
$P_s=0$ and $q_r=0$. By inspection of Eq. (\ref{loadv3}) we then
obtain the mobility $\mu=1/(\zeta_p+\zeta_v)$, as discussed in
Sec. IIIb. (ii) Similarly, in the strained state S, the protein
friction $\zeta(t)$ is locked onto $\zeta_v$, $P_s=1$, $P_r=0$,
and $q_s=0$, and we obtain the mobility $\mu=1/2\zeta_v$. (iii)
Finally, in the case $\zeta_p=\zeta_v$, we immediately find
$\mu=1/2\zeta_v$, as is also evident from the model equations
(\ref{lan1},\ref{lan2}).
Interpolating between the limiting cases (i) and (ii) above, we
introduce the average mobility according to
\begin{equation}
\label{muav}
\mu_{\text{av}}=P_r\frac{1}{\zeta_p+\zeta_v}+P_s\frac{1}{2\zeta_v}=
\frac{\zeta_v+P_r\zeta_v+P_s\zeta_p}{2\zeta_v(\zeta_p+\zeta_v)}.
\end{equation}
and the mobility in Eq. (\ref{loadv3}) takes the form
\begin{equation}
\label{loadv4} \mu=\mu_{\text{av}}\cdot \frac{1+
(2q_s+(1+\frac{\zeta_v}{\zeta_p})q_r)
\frac{P_r\zeta_v+P_s\zeta_p}{P_r\zeta_v+P_s\zeta_p+\zeta_v}} {
1+q_s+q_r}.
\end{equation}

Inserting the characteristic biological numbers of Mogilner et
al.\cite{Mogilner98}: $\zeta_p/\zeta_v = 50$, $q_r=0.2$,
$q_s=2\cdot 10^{-3}$, and $P_r\sim P_s\sim 0.5$, we obtain the
average mobility $\mu_{\text{av}}\approx 2.6\times 10^5$ nm/(s
pN), while the correction factor in Eq. (\ref{loadv4}) is $0.998$,
i.e., very close to 1. Hence, this gives rise to the ratio
\begin{eqnarray}
\frac{\mu}{\mu_{\text{M}}}\approx 13,
\end{eqnarray}
in comparison with the value estimated in Ref.~\cite{Mogilner98}:
\begin{eqnarray}
\label{muM}
\mu_{\text{M}}=\frac{1}{\zeta_p}.
\end{eqnarray}
The origin of the discrepancy between the present result and that
of Ref.~\cite{Mogilner98} is the slight difference between the
models. In the strained state Ref.~\cite{Mogilner98} operates with
two characteristic times, that of $S\rightarrow R$ conversion,
i.e., the residence time Eq. (\ref{residence}) $\langle
t\rangle_s=1/g_r$, and the time of the restoration of bonds
between the motor working head and the groove, which is much
smaller. In the present model the two times are assumed equal,
corresponding to  the assumption that the spring relaxation
$S\rightarrow R$ is initiated when the working head becomes
attached to the groove again. From a physical point of view this
is equally possible, but implies that the motor spends much
shorter times in state S than in state R, or $g_r\gg g_s$, i.e.,
$P_r\sim 1$ and $P_s\sim 0$, and the motor becomes much more
volatile to the local force during these periods.

\begin{figure}
\includegraphics[width=0.5\hsize,angle=-90]{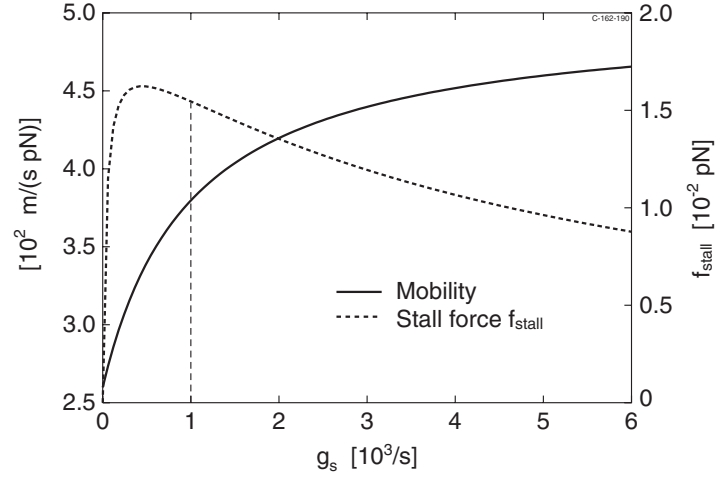}
\caption{ Mobility $\mu$ and stall force $f_{\rm stall}$ as a
function of the hydrolysis rate $g_s$. The vertical line marks the
value from Ref.~\protect\cite{Mogilner98}. Note the different
scales.} \label{fig4}
\end{figure}

\begin{figure}
\includegraphics[width=0.5\hsize,angle=-90]{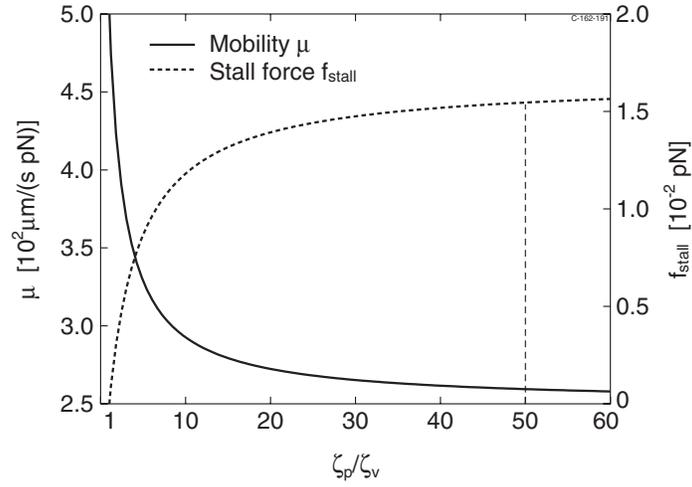}
\caption{Mobility $\mu$ and stall force $f_{\rm stall}$ as a
function of the protein friction $\zeta_p$ (model value $5.0\cdot
10^{-5}$ nm/(s pN)).} \label{fig5}
\end{figure}

\begin{figure}
\includegraphics[width=0.5\hsize]{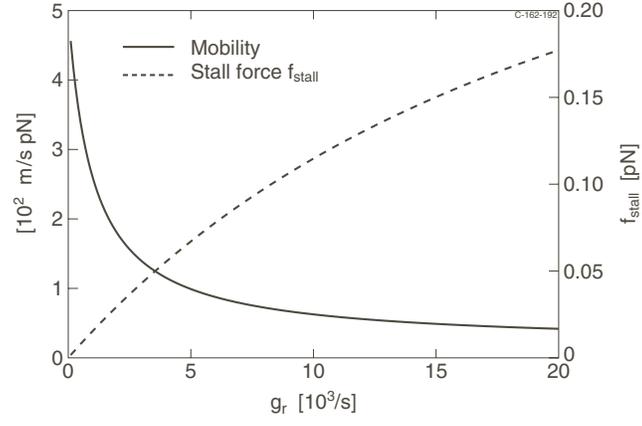}
\caption{Mobility $\mu$ and stall force $f_{\text{stall}}$ as a
function of the rate of relaxation $g_r$. We observe that for
large relaxation rate the stall force enters the biological
range.} \label{fig6}
\end{figure}

From the mobility $\mu$ and the zero-load velocity $\langle
v_x^0\rangle$, we obtain the stall force
\begin{equation}
\label{fstall} f_{\text{stall}}=\frac{\langle v_x^0\rangle}{\mu}.
\end{equation}
In Figs.~\ref{fig4} and \ref{fig5} we have on the same plots
depicted the values of the mobility $\mu$ and the stall force
$f_{\text{stall}}$ versus the hydrolysis rate $g_s$ and the
protein friction $\zeta_p$. We note that the stall force exhibits
a maximum as function of $g_s$ close to the parameter values
chosen in our calculations, whereas the mobility is close to
saturation. Similarly, as function of $\zeta_p$, the stall force
is close to its maximum value, whereas the mobility does not
change much within the chosen plot range (note that the ordinate
does not reach the origin). In general, we observe that due to the
particular dependence of $\mu$ on the model parameters, its value
varies relatively weakly within large intervals for the individual
parameter values. In Fig. ~\ref{fig6} we have depicted the
mobility and stall force as functions of the rate of relaxation
$g_r$. For large $g_r$ we obtain a stall force of the order pN,
which is in the biological range.

\section{\label{conclusion}Summary and Conclusion}
In this paper we have by analytical means solved a molecular motor
model proposed by Mogilner et al. \cite{Mogilner98}. This model
represents a robotic motor solely based on an effective static
friction interaction between motor and its support (track). From
the underlying Langevin equations, which represent the
synchronized dichotomous processes of friction, effective spring
constant and distance between the motor heads, we obtain explicit
expressions for the load-free motor velocity, the mobility of the
motor, and the stall force. Whereas the result for the load-free
velocity produces a typical motor velocity of several $\mu$m/s for
physiologically reasonable parameters, the exact solution
overestimates the mobility, leading to a value for the stall
force, which is roughly two orders of magnitude smaller than
physiological values and significantly smaller than the
simulations results reported in Ref.~\cite{Mogilner98}. This
variance is associated with a difference in the stochastic
dynamics underlying the analysis and the dynamical processes
implied in the numerical simulation.

A likely explanation relies on the essential feature of the model,
which is the decoupling of the dynamics of the motor protein-rail
biopolymer interaction (chemical bonds forming and breaking) into
a fast, detached process during energy consumption, and a slow
relaxing process in one mechanochemical motor cycle. This purely
stochastic picture leads to situations in which the motor detaches
frequently, before its relaxing step is finished, and therefore
the sub-cycles, which actually lend themselves to propulsion, are
interrupted. Obviously, this leads to the underestimation of the
stall force. In a real system, the fact that chemical bonds are
established ensures that a full propulsion sub-cycle can be
completed before dissociation takes place for the next loading of
the internal motor `spring' in parallel to hydrolysis. In
comparison to the ratchet models in which the motor properties are
represented by fluctuating between two different, periodic
potentials, it appears that the latter rely on fewer parameters,
and therefore their stall force can be adjusted better to actually
observed values.

We finally should like to emphasize that the obtained exact
results allow for an exact and detailed study of the dependence of
the motor characteristics on the various parameters without
invoking numerical simulations. Additional features such as the
low likelihood for detaching from the rail during the
forward-motion, could be incorporated into the model and still be
solved explicitly, using the solution schemes developed here. We
therefore believe that this study leads to a better understanding
of molecular motor models.

\acknowledgments We should like to thank John Hertz for very
constructive discussions.


\newpage

\vspace{2cm}

\end{document}